# Predicting microsatellite instability and key biomarkers in colorectal cancer from H&E-stained images: Achieving SOTA predictive performance with fewer data using Swin Transformer

Bangwei Guo[1], Xingyu Li[2], Jitendra Jonnagaddala[3*], Hong Zhang[2*], Xu Steven Xu[4*]

[1] School of Data Science, University of Science and Technology of China

[2] Department of Statistics and Finance, School of Management, University of Science and Technology of China

[3] School of Population Health, UNSW Sydney, Kensington, NSW, Australia

[4] Data Science/Translational Research, Genmab Inc., Princeton, New Jersey, USA

*Corresponding author. jitendra.jonnagaddala@unsw.edu.au (Jitendra Jonnagaddala), zhangh@ustc.edu.cn (Hong Zhang), sxu@genmab.com (Xu Steven Xu)

## Abstract

Artificial intelligence (AI) models have been developed to predict clinically relevant biomarkers for colorectal cancer (CRC), including microsatellite instability (MSI). However, existing deep-learning networks are data-hungry and require large training datasets, which are often lacking in the medical domain. In this study, based on the latest Hierarchical Vision Transformer using Shifted Windows (Swin-T), we developed an efficient workflow for biomarkers in CRC (MSI, hypermutation, chromosomal instability, CpG island methylator phenotype, *BRAF,* and *TP53* mutation) that required relatively small datasets, but achieved a state-of-the-art (SOTA) predictive performance. Our Swin-T workflow substantially outperformed published models in an intra-study cross-validation experiment using the TCGA-CRC-DX dataset (N = 462). It also demonstrated excellent generalizability in cross-study external validation and delivered a SOTA AUROC of 0.90 for MSI, using the MCO dataset for training (N = 1065) and the TCGA-CRC-DX for testing. A similar performance (AUROC = 0.91) was achieved by Echle et al., using ~8000 training samples (ResNet18) on the same testing dataset. Swin-T was extremely efficient when using small training datasets and exhibited robust predictive performance with 200–500 training samples. These data indicate that Swin-T could be 5–10 times more efficient than existing algorithms for MSI based on ResNet18 and ShuffleNet. Furthermore, the Swin-T models showed promise as pre-screening tests for MSI status and BRAF mutation status, which could exclude and reduce the samples before subsequent standard testing in a cascading diagnostic workflow, to allow a reduction in turnaround time and costs.

# Introduction

Artificial intelligence (AI) and deep-learning models using hematoxylin and eosin (H&E)-stained histology whole-slide images (WSIs) have been developed to predict clinically relevant molecular biomarkers for colorectal cancer (CRC), such as microsatellite instability (MSI)[1-3], genetic mutations[4-6], and molecular subtypes[6,7]. In particular, MSI prediction is of great clinical utility since it is the only approved biomarker to select patients for immune checkpoint inhibitors[8]. The US Food and Drug Administration (FDA) granted accelerated approval for anti-PD1 antibodies (such as pembrolizumab and nivolumab) for the treatment of MSI-high or mismatch repair deficient (dMMR) cancers, including CRC (the FDA's first tissue/site-agnostic approval)[9,10].

Kather et al. developed the first fully automated deep learning (DL) model for predicting MSI/dMMR status in CRC in 2019[1]. Since then, multiple models have been published[1-4,6,11-16]. Recently, the vision transformer (ViT)[17] has emerged as a new modeling framework in the field of computer vision and has shown great potential to replace the convolutional neural network (CNN)[14,17], which has been the backbone of the vast majority of DL models in digital pathology (including MSI models). Earlier versions of ViT still require very large datasets to achieve a performance comparable to that of CNN models[17,18].

In the medical imaging domain, datasets are usually limited and often accompanied by weak (slide-level) labels[19]. In addition, attention-based DL models such as ViT are complex and data-hungry in nature[19,20]. These challenges pose a significant barrier to the development and application of DL models in digital pathology. For example, the current state-of-the-art (SOTA) performance for predicting MSI status was achieved using extremely large pooled datasets from different studies (N > 8,000 WSIs)[11]. Therefore, achieving clinical-grade prediction of MSI status and other key biomarkers for CRC using limited data remains an active research question. The latest hierarchical vision transformer using shifted windows (Swin Transformer; Swin-T hereafter) reduces computational complexity and can flexibly process pictures with different scales[18]. Therefore, Swin-T may have the potential to circumvent the limitations of small datasets in medical image studies.

In this study, we aimed to develop an efficient workflow using Swin-T that can use relatively small training datasets but achieves the best SOTA predictive performance for MSI status and other key biomarkers in CRC (BRAF mutation, TP53 mutation, CpG island methylator phenotype (CIMP), hypermutation, and chromosomal instability), using H&E-stained images of colorectal tumors.

## Method

The workflow for processing the Whole Slide Images (WSIs) and modeling the data is illustrated in **Figure 1**. In this study, we developed a novel Swin-T-based deep learning pipeline for predicting key biomarkers in CRC patients, including MSI status. This pipeline included two Swin-T models: a tissue classifier to detect tumor tissues and a biomarker classifier to predict the binary biomarker status.

**Imaging and clinical data**

Two international CRC datasets were analyzed in this study. The Molecular and Cellular Oncology (MCO) dataset is a collection of any stage patients who underwent curative resection for colorectal cancer between 1994 and 2010 in New South Wales, Australia. The TCGA dataset ('The Cancer Genome Atlas', publicly available at https://portal.gdc.cancer.gov/, USA) is a multicentric collection of tissue specimens, which include tumors of all stages in the TCGA-COAD and TCGA-READ datasets.

Anonymized H&E stained WSIs were collected from two datasets with matched genomic data. For the MCO dataset, a small number of patients (n = 73) were excluded because of the absence of tumor tiles and lack of molecular information. After excluding, 1,065 WSIs from 1,065 patients in the MCO dataset were used in this study. The ground truth labels of the MCO dataset were available for MSI status (157 microsatellites unstable and 908 microsatellites stable), *BRAF* mutation (117 mutants and 909 wild-types), and CIMP (153 CIMP high and 211 CIMP low). The TCGA-CRC-DX dataset, which has been used in multiple previously published studies for MSI prediction[4,11,12,14], includes 502 WSIs of primary colorectal tumors from 499 patients. A summary of the ground-truth labels in the TCGA-CRC-DX dataset for different biomarkers (hypermutation, microsatellite instability, chromosomal instability, CIMP, *BRAF*, and *TP53* mutations) is available in Bilal et al[4].

**Data preprocessing**

Scanned WSIs were downloaded in SVS format. All the WSIs were tessellated into small image tiles of 512 × 512 pixels at a resolution of 0.5 μm. No manual annotations of the tumor tissue were used. The image tiles were color-normalized using Macenko's method[21] to reduce the color bias and improve classifier performance and were subsequently resized to 224 × 224 pixels to serve as the input of the network. Image tiles containing background or blurry images were automatically removed from the dataset during this process, using the detected edge quantity (Canny edge detection in Python's OpenCV package) (https://github.com/KatherLab/preProcessing).

A Swin-T tissue classifier was trained to detect and select tiles with tumor tissue using a publicly available dataset, NCT-CRC-HE-100K, which consists of nine types of CRC tissue images collected by Kather et al[22]. Up to 500 tumor tiles were randomly selected per patient and used for all subsequent steps. All selected tiles inherit the label of the corresponding patient. Therefore, all models were trained using only slide-level labels.

**Training strategy for deep learning models**

A stepwise strategy was adopted during the model development.

*Pre-training: a Swin-T tissue classifier*

First, a Swin-T model was pre-trained to develop a multiclass tissue classifier. The tissue classifier was trained and tested using two publicly available pathologist-annotated datasets (NCT-CRC-HE-100K and CRC-VAL-HE-7K) from Kather et al. These datasets consist of CRC image tiles of nine tissue types: adipose tissue (ADI), background (BACK), debris (DEB), lymphocytes (LYM), mucus (MUC), smooth muscle (MUS), normal colon mucosa (NORM), cancer-associated stroma (STR), and colorectal adenocarcinoma epithelium (TUM) [23]. The NCT-HE-100K dataset contains 100,000 image tiles (TUM = 14,317 and non-tumor = 85,683). All image patches were 224 × 224 pixels at 20X magnification. During the pre-training process, the model was trained for 10 epochs using a fixed learning rate of 0.001. The Adam optimizer and cross-entropy loss functions were used. After pre-training was completed, the parameter weights of the backbone part of the pre-trained model were saved. The Swin-T tissue classifier achieved a high overall accuracy (96.3%) and tumor detection accuracy (98%) of the tissue segmentation model on the unseen test set, CRC-VAL-HE-7K, which contained 7,180 image tiles (TUM = 1,223 and non-tumor = 6,957) (Figure 1).

### *Fine-tuning: Swin-T biomarker models*

Subsequently, the pre-trained Swin-T model (tissue classifier) was fine-tuned for the binary classification of key CRC biomarkers at the patient (slide) level (such as microsatellite unstable *vs*. microsatellite stable for MSI, high mutation density *vs*. low mutation density for hypermutation, chromosomal instability *vs*. genomic stability for chromosomal instability, CIMP-high *vs*. CIMP-low for CIMP, mutant vs. wild type for *BRAF,* and *TP53* mutation). The linear project layers of the saved pre-trained Swin-T model for tissue classification were replaced by new linear layers to accommodate the prediction of the binary classification of CRC markers. The new model was fine-tuned for 20 epochs in the binary classification task for each binary CRC biomarker, using a decaying learning rate policy (the learning rate at the $n$th epoch was $0.0001/n$). Similarly, Adam optimizer and weighted cross-entropy loss function were used during fine-tuning. The average value of the predictive scores of all tiles for each WSI was calculated as the score for the corresponding molecular status of the whole-slide image.

**Experimental setup**

Three experiments were conducted to evaluate the performance of our Swin-T model. First,

we compared the predictive performance of the Swin-T models with that of state-of-the-art models for predicting six CRC biomarkers (hypermutation, microsatellite instability, chromosomal instability, CIMP, *BRAF*, and *TP53* mutations) from two recent publications[4,6] using intra-study cross-validation. For all six biomarkers, we used the same training-to-test dataset split of the TCGA-CRC-DX dataset for four-fold cross-validation, used and published by Kather et al.[6] or/and Bilal et al.[4] The split-match of the TCGA-CRC-DX cohort facilitated the comparison of model performance with previous publications. In cross-validation, one-fold of the training set was used as the validation set to select the best-performing model, which was saved for testing on the unseen test fold.

Second, the predictive performance of Swin-T models was compared with state-of-the-art models from recent publications using cross-study external validation for predicting microsatellite instability and *BRAF* mutation[11,14]. For the external validation experiments, the pre-trained Swin-T model was fine-tuned using the MCO cohort to develop models for predicting MSI status and *BRAF* mutations. In addition, a model for predicting the CIMP status was developed, as the molecular data for these three biomarkers are available in both TCGA and MCO cohorts. In this experiment, the fine-tuned models were tested externally on the unseen TCGA-CRC-DX cohort to facilitate comparison with publications in which external validation was performed on the same cohort.

Finally, to understand the impact of the sample size of the training data on the performance of the Swin-T models, we randomly selected 25, 50, and 75% of the MCO data and trained additional models for the prediction of MSI status using the same strategy as that used during the training with all MCO data. Similarly, the external validation performance of the TCGA-CRC-DX cohort was compared for different sample sizes for the training data.

**Statistical analyses**

The predictive performance of the deep learning models was evaluated using different statistical metrics. For the intra-study cross-validation, the mean values of the area under the receiver operating curve (AUROC) and the area under the precision-recall curve (AUPRC) were used by averaging four-fold. The AUPRC values were compared to account for the imbalanced data commonly observed in clinical studies. The standard deviations were calculated. For external validation, AUROC and AUPRC were used to compare the models. The bootstrap method (1,000 ×) was applied to calculate 95% confidence intervals (CI) for the external validation of AUROC and AUPRC. To evaluate the feasibility of the models as diagnostic tools, different classification thresholds were set to calculate various statistical metrics to evaluate diagnostic performance based on external validation experiments. The thresholds included a cutoff at fixed 95% sensitivity and a cutoff at fixed values of 0.25, 0.50, and 0.75. The sensitivity, specificity, positive predictive value, negative predictive value, true negative fraction, false-negative fraction, and F1-score were used to evaluate the diagnostic values of the AI models. In addition, the interpretability of the Swin-T models was explored using visualization technology with Python package pytorch_grad_cam (https://github.com/jacobgil/pytorch-grad-cam). The Grad-CAM method[24] was applied to visualize the activation feature map of the Swin-T model and interpret the model outcome.

# Results

**1. Swin-T provides an excellent predictive performance**

### Intra-study cross-validation using the TCGA-CRC-DX dataset

In this experiment, six molecular biomarkers (MSI, hypermutation, CIMP, CING, *BRAF* mutation, and *TP53* mutation) were predicted using Swin-T. Kather et al.[6] and Bilal et al.[4] used a TCGA-CRC-DX dataset to develop and evaluate their deep-learning models for MSI status and other key biomarkers for CRC, via intra-study cross-validation. To facilitate comparison with their existing models, we used the same patient cohort and the same four-fold splits of the TCGA dataset published by Bilal et al.[4]

For predicting the high microsatellite instability status, our Swin-T achieved a mean validation AUROC of 0.91 $\pm$ 0.03 (mean $\pm$ standard deviation (SD)), which represents approximately 6 and 23% improvement over recently published AUROC values on the same dataset, 0.86 from Bilal et al.[4] and 0.74 from Kather et al.[12], respectively (**Table 1**). For predicting the hypermutation status, Swin-T also outperformed the models developed by Kather et al.[6] and Bilal et al.[4] on the same dataset, and significantly improved the predictive performance. The AUROC based on Swin-T was 0.85 $\pm$ 0.02, compared with 0.81 and 0.71 reported by Bilal et al.[4] and Kather et al.[6], respectively. Moreover, Swin-T exhibited similar performance to that of Bilal et al.[4] for predicting *TP53* mutation status (AUROC = 0.73) but was significantly higher than that obtained by Kather et al.[6] (AUROC = 0.64). Furthermore, although our mean cross-validation AUROC values for predicting chromosomal instability (vs. genomic stability), *BRAF* mutation status, and high CIMP status were slightly lower than those reported by Bilal et al.[4], the difference was only about 1–2%.

Overall, for a fixed sample size of training set (the same TCGA-CRC-DX dataset), Swin-T significantly improved the prediction of MSI and hypermutation status in intra-study cross-validation, over published models. Moreover, Swin-T achieved the SOTA performance in predicting *TP53* mutations. Furthermore, Swin-T provided similar or higher AUPRCs for MSI, hypermutation, CIMP, and BRAF mutation status compared with the models reported by Bilal et al.[4], suggesting that the Swin-T models better predict positive results (MSI-high, high mutation density, CIMP high, and BRAF mutants) with a lower false positive rate.

*Cross-study external validation using TCGA-CRC-DX dataset*

The generalizability of a model is often evaluated through cross-study external validations. The TCGA-CRC-DX dataset has been used for the external validation of multiple AI models for predicting MSI status. In this experiment, we trained the Swin-T model using the MCO dataset (N = 1065) and externally validated the model using the TCGA-CRC-DX dataset to compare the model performance. The Swin-T model yielded an excellent mean external validation AUROC of 0.904 (95% confidence interval (CI): 0.849 – 0.952; **Table 2**). In comparison, Echle et al.[11] trained a CNN model using a combined dataset from multiple large international studies (N = 7917) and achieved a similar mean external-validation AUROC of 0.91 (95% CI: 0.87–0.95). The model trained using ShuffleNet and a similar size of training data (N ranging from approximately 1,000–2,000) only produced AUROC values from 0.72 to 0.77 (**Table 2**)[12]. Therefore, Swin-T achieved similar SOTA generalizability in external validation compared with the most recently published model trained on a large, pooled dataset.

Swin-T also displayed similar SOTA performance to previously published methods for predicting *BRAF* mutation status (AUROC: 0.80 *vs* 0.81) in external validation using the TCGA-CRC-DX dataset. It is worth noting that Swin-T produced substantially better AUPRC values compared with previous publications for predicting both MSI status (AUROC: 0.66 vs. 0.62) and

*BRAF* mutation (AUROC: 0.35 vs 0.33).

Furthermore, the Swin-T architecture demonstrated a great potential for minimizing overfitting, which is often observed in deep-learning modeling, and produced very similar predictive performance between the training and external validation datasets. In the four-fold cross-validation experiment of the MCO dataset for predicting MSI status, Swin-T achieved a mean AUROC value of $0.926 \pm 0.055$, compared to 0.904 (95% CI: 0.849–0.952) in the external validation dataset TCGA-CRC-DX. A similar pattern was observed for BRAF (AUROC: 0.88 vs. 0.80) and CIMP (AUROC: 0.766 vs. 0.759) (Figure 3).

## 2. Swin-T models as diagnostic tools

Based on external validation using the TCGA-CRC-DX cohort, we also evaluated the feasibility of using Swin-T models as diagnostic tools for MSI/dMMR status, *BRAF* mutation, and CIMP status, based on routine digitized H&E-stained tissue slides of CRC. Computer-based AI systems are often positioned as pre-screening tools before the gold-standard confirmatory tests[11]. Therefore, the clinical utility of these pre-screening tools is primarily to minimize false negative predictions but exclude as many true negative samples as possible from the subsequent confirmatory test runs.

### *Pre-screening for MSI status*

For diagnostic purposes, a cutoff value is required to determine the diagnostic outcome. **Table 3** shows that for a cutoff that can provide 95% sensitivity for detecting MSI-high patients (cutoff = 0.16), the negative predictive value was 98% and the false negative fraction was 0.7%.

Meanwhile, the true negative fraction was 34% with this cut-off, which implies that 34% of patients can be safely excluded from confirmatory tests in clinical settings. Similar results were reported by Echle et al.[11] for MSI detection. When a fixed cutoff of 0.25, the sensitivity was reduced slightly to 92%. The Negative Predictive Value (NPV) remained almost the same at 98%, whereas the False-Negative Fraction (FNF) increased slightly to 1.2%. However, with a cut-off of 0.25, 55.4% of the patients were excluded from the confirmatory tests. These results confirmed previous reports that AI models can serve as pre-screening purposes for MSI status[8,11,12].

### *Pre-screening for BRAF mutation*

For the Swin-T model for *BRAF* mutations, the cut-off for 95% sensitivity was 0.17. At such a cutoff, the NPV was 96%, the FNF was 0.4%, and the True Negative Fraction (TNF) was 10%, suggesting that 10% of *BRAF* WT patients would be safely excluded from the gold-standard confirmatory testing for *BRAF* mutation. However, if we used a fixed cutoff of 0.25, 27.2% of patients could be correctly determined as BRAF WT, whereas the false negatives (patients incorrectly determined as *BRAF* WT) remained low (0.8%). Of the predicted *BRAF* WT, 96% were *BRAF* WT at a cut-off of 0.25. The Swin-T model for *BRAF* mutations exhibited potential as a pre-screening AI diagnostic tool for *BRAF* mutations.

### *Pre-screening for CIMP status*

A cutoff of 0.1 can provide 95% sensitivity for predicting CIMP status. At this cutoff, 1.3% CIMP high would be incorrectly identified as CIMP-low, whereas 16.2% true CIMP-low can be excluded from subsequent confirmatory molecular testing. However, when the cut-off was increased to 0.25, the false negatives substantially increased to approximately 5%. Therefore, the performance of the current Swin-T model for the CIMP may not be optimal as a diagnostic tool.

## 3. Swin-T exhibits robustness with small training data

As expected, the larger the training datasets, the higher the predictive performance (**Figure 5**). However, the Swin-T model maintained a cross-study AUROC of 0.864 (on the external validation dataset TCGA-CRC-DX) for predicting microsatellite instability when only 50% of MCO data (N ~500) were used in training. In the scenario using only 25% MCO dataset for training (N ~250), the Swin-T model could still produce an impressive AUROC value of 0.806 for predicting the MSI status.

## 4. Swin-T models improve interpretability

The tile-level interpretation of deep-learning models for MSI has been popular and is often provided in the current literature[1,3,8,12]. We visualized the highest-scoring tiles of patients with the highest WSI scores (**Figure 6a**). For Swin-T, we also drew a cam map for a tile (**Figure 6b**), which is a heatmap highlighting the areas (cells) that have greater attention weights within a high-resolution tile. The brighter the color of an area, the higher the attention scores assigned to this area by the model. **Figure 6b** shows that within the tiles of MSI samples, the model focuses on the microenvironment around the tumor cells rather than the tumor cells themselves, including mucus, infiltrating lymphocytes, and tumor stroma. However, within the tiles predicted from the microsatellite stability (MSS) sample, the model mainly focused on tumor tissues.

These findings were consistent with the pathological characteristics of MSI-high or MSS samples[25,26], suggesting that coupling Swin-T with the Grad-CAM algorithm[24] could further improve the interpretability of the model and detect the areas or cells within tiles that have great potential to aid pathologists in identifying novel image features associated with the classification of a biomarker.

## Discussion

In this study, we developed a novel deep-learning framework based on a Swin-T backbone network to predict MSI status and other key biomarkers for CRC. The Swin-T backbone represents the most advanced, start-of-the-art vision transformer network architecture and has achieved outstanding performance in many computer-vision tasks, outperforming many popular de facto standard networks such as ResNet and EfficientNet, as well as early versions of vision transformers (ViT)[17,18]. It was proven that the Swin-T can replace the classic CNN architecture and become a common backbone in the field of computer vision[18]. However, to our knowledge, despite having achieved great success in common computer vision tasks, our work represents the first attempt to evaluate the performance of Swin-T in digital pathology and as a backbone network to further improve the predictive performance of MSI and biomarkers for molecular pathways in CRC.

We demonstrated that novel Swin-T-based backbone networks have great utility in digital pathology and can improve the predictive performance for microsatellite instability and other key biomarkers in CRC. To facilitate comparison with previously published models, the same dataset (TCGA-CRC-DX) and training-to-test split of the dataset from previous publications were used.

In an intra-study cross-validation experiment, Swin-T substantially outperformed models by Bilal et al.[4] and Kather et al.[12] for the prediction of microsatellite instability and hypermutation status. In addition, Swin-T achieved a similar SOTA performance for predicting *TP53* mutation status compared with that reported by Bilal et al.[4] Similar mean cross-validation AUROC values were also obtained for predicting chromosomal instability, *BRAF* mutation status, and high CIMP status compared with the current literature [4]. Swin-T models also exhibited similar or higher AUPRCs for MSI, hypermutation, CIMP, and *BRAF* mutation status compared with previously published state-of-the-art computational algorithms[4], indicative of greater power for handling imbalanced data, often seen in clinical studies.

It is well known that deep-learning models perform better with more available training data. This phenomenon has been observed in prediction models developed for MSI/dMMR status in CRC[12]. Recently, Echle et al. trained a model using pooled data from nine patient cohorts of 8,343 patients across different countries and ethnicities, and achieved SOTA external prediction performance with an AUC of 0.91, using the TCGA-CRC-DX cohort as the external validation dataset[11]. However, with smaller training data (QUASAR: N = 1,016; DACHS: N = 2,013; NLCS: N = 2,197), Echle et al. (the same research group) obtained an AUROC of 0.72–0.77, with the same unseen external validation cohort[12]. Swin-T demonstrated excellent generalizability in cross-study external validation using the same TCGA-CRC-DX dataset and delivered a SOTA AUROC of 0.904 using a relatively smaller training dataset (MCO, N = 1,065), similar to what was achieved by Echle et al. using ~8,000 samples (ResNet18). Our additional experiment revealed that Swin-T was extremely efficient when using small training datasets. Using ~250 samples for training, the Swin-T model still managed to produce better predictive performance than the model by Echle et al. using ShuffleNet and training data of 1,000–2,000 samples[12]. These results suggest that our MSI model based on Swin-T may be 5–10 times more efficient than current state-of-the-art MSI algorithms based on ResNet18 and ShuffleNet.

Biomarker testing plays a critical role in the treatment of CRC patients. Importantly, immunotherapies, such as pembrolizumab and nivolumab, were approved by health authorities to treat CRC patients with MSI-High[9,10]. The current clinical gold-standard testing for MSI is based on immunohistochemistry, which has a sensitivity of 94% and specificity of 88%[27,28]. The motivation to develop AI-based models is primarily to replace current lab-based testing, reduce the turnaround time, and save costs. Unfortunately, thus far, no digital AI models for MSI have consistently achieved this performance threshold, including the most recent model developed by Echle et al[11]. Therefore, it is proposed to implement the current SOTA MSI models as a pre-screening test, which primarily excludes and reduces the samples before the subsequent conventional IHC testing[11] Our Swin-T model for MSI status displayed a diagnostic performance similar to that of the SOTA model developed by Echle et al. (trained using data from approximately 8, 000 CRC patients). In addition, the Swin-T model for *BRAF* mutation status showed promise as a pre-screening test, although further improvement might be required. These results demonstrate the potential of this Swin-T-based AI system as an important component in a cascading diagnostic workflow (pre-screening + gold-standard testing) for MSI and *BRAF* mutation status, which are important for patient selection in clinical trials and treatment guidance for immune checkpoint inhibitors and combinations of BRAF inhibitors/anti-epidermal growth factor receptor therapies, respectively.

**Table 1. Comparison of predictive performance of Swin-T for key CRC biomarkers with published models using intra-study cross validation on TCGA-CRC-DX**

| Biomarker | AUROC | | | AUPRC | |
|---|---|---|---|---|---|
| | Swin-T | Kather et al.[6] | Bilal et al.[4] (IDaRS) | Swin-T | Bilal et al.[4] (IDaRS) |
| Four-fold cross-validation in the TCGA-CRC-DX cohort | | | | | |
| Microsatellite instability vs stability | **0.91± 0.02** | 0.74 | 0.86± 0.03 | **0.66± 0.09** | 0.62± 0.10 |
| High vs low mutation density | **0.85± 0.03** | 0.71 | 0.81±0.04 | **0.58 ± 0.05** | 0.57± 0.09 |
| Chromosomal instability vs genomic stability | 0.82± 0.04 | 0.73 | **0.83± 0.02** | 0.90± 0.03 | **0.92± 0.01** |
| CIMP-high vs CIMP-low | 0.77± 0.06 | ND | **0.79± 0.05** | **0.60± 0.15** | 0.51± 0.05 |
| BRAF | 0.77± 0.02 | 0.66 | **0.79± 0.01** | **0.35± 0.11** | 0.33± 0.05 |
| TP53 | **0.73± 0.02** | 0.64 | **0.73± 0.02** | 0.75± 0.02 | **0.78± 0.04** |

ND: Not Done. IDaRS: Iterative Draw and Rank Sampling

**Table 2. Comparison of predictive performance of MCO-trained Swin-T for MSI status and BRAF mutation with published models using external validation on TCGA-CRC-DX.**

| Network | Training Dataset | Number of training samples | AUROC (95% CI) | AUPRC (95% CI) |
|---|---|---|---|---|
| **MSI** | | | | |
| Swin-T(Ours) | MCO | 1,065 | **0.90 (0.85-0.95)** | 0.718 (0.605-0.820) |
| VIT[14] (2022) | DACHS | 2,069 | 0.89 (0.83-0.93) | 0.672 (0.558-0.769) |
| ResNet18[11] (2022) | Pooled International datasets | 7,917 | **0.91 (0.87-0.95)** | … |
| ShuffleNet[12] (2020) | QUASAR | 1,016 | 0.76 (0.70-0.79) | … |
| ShuffleNet[12] (2020) | DACHS | 2,013 | 0.77 (0.73-0.79) | … |
| ShuffleNet[12] (2020) | NLCS | 2,197 | 0.72 (0.71-0.78) | … |
| **BRAF** | | | | |
| Swin-T(Ours) | MCO | 1,026 | **0.80 (0.74-0.87)** | 0.392 (0.279-0.541) |
| EfficientNet[14] (2022) | DACHS | 2,069 | **0.81 (0.75-0.86)** | 0.360 (0.253-0.487) |

**Table 3. Statistics results using different thresholds of external validation of predictions for the MSI, BRAF and CIMP status in TCGA-CRC-DX cohort.**

| Biomarker | Threshold | Sensitivity | Specificity | PPV | NPV | TNF | FNF | F1 Score |
|---|---|---|---|---|---|---|---|---|
| MSI | 0.16 | 0.95 | 0.402 | 0.209 | 0.980 | 0.340 | 0.007 | 0.343 |
| | 0.25 | 0.918 | 0.647 | 0.304 | 0.979 | 0.554 | 0.012 | 0.457 |
| | 0.50 | 0.721 | 0.939 | 0.667 | 0.953 | 0.804 | 0.040 | 0.693 |
| | 0.75 | 0.213 | 0.997 | 0.928 | 0.883 | 0.853 | 0.113 | 0.346 |
| BRAF | 0.17 | 0.950 | 0.114 | 0.123 | 0.959 | 0.095 | 0.004 | 0.218 |
| | 0.25 | 0.930 | 0.307 | 0.148 | 0.971 | 0.272 | 0.008 | 0.255 |
| | 0.50 | 0.649 | 0.827 | 0.327 | 0.948 | 0.732 | 0.040 | 0.435 |
| | 0.75 | 0.263 | 0.977 | 0.6 | 0.911 | 0.865 | 0.085 | 0.366 |
| CIMP | 0.10 | 0.950 | 0.201 | 0.263 | 0.927 | 0.162 | 0.013 | 0.411 |
| | 0.25 | 0.796 | 0.536 | 0.339 | 0.898 | 0.413 | 0.047 | 0.475 |
| | 0.50 | 0.556 | 0.845 | 0.517 | 0.864 | 0.651 | 0.102 | 0.536 |
| | 0.75 | 0.278 | 0.945 | 0.6 | 0.814 | 0.728 | 0.166 | 0.380 |

Statistics describe the different thresholds when the network is trained on MCO cohorts and tested on TCGA-CRC-DX cohort. PPV: Positive Predictive Value. NPV: Negative Predictive Value. TNF: True-Negative Fraction (Rule-out). FNF: False-Negative Fraction.

**Figures**

**Figure 1: The workflow of the data preprocessing and the training process of the deep-learning model. (a)** Tiles images of NCT-CRC-HE-100K are downloaded from the publicly available website (https://zenodo.org/record/1214456) to pre-train a tissue classifier based on Swin-T. The classifier has excellent performance of classifying tissues (overall accuracy = 96.3%) and detecting tumor patches (accuracy = 98%) in an external dataset: CRC-VAL-HE-7K. **(b)** Whole-Slide images in the SVS format of the MCO dataset and TCGA dataset are preprocessed to tessellate into non-overlapping patches with a size of 512 × 512 pixels. These tiles are then resized to the smaller 224×224 pixels tiles and color normalized. The tumor patches are selected. **(c)** For each patient, up to 500 tiles are randomly sampled for subsequent experiments. The pre-trained tissue classifier model in (a) is then fine-tuned to predict biomarker status of each tile. The probability values of the tiles are averaged to derive the prediction at the patient level. The performance of the models is evaluated in two separate experiments: an intra-cohort four-fold cross-validation and an inter-cohort external validation.

**Figure 2: Predictive performance of four-fold cross-validation of Swin-T based prediction of colorectal cancer biomarkers in the TCGA-CRC-DX cohort.** AUROC plots for prediction of hypermutation (HM), microsatellite instability (MSI), chromosomal instability (CING), CpG island methylator phenotype (CIMP), BRAF mutation status, and TP53 mutation status. The True Positive Rate represents sensitivity and the False Positive Rate represents 1–specificity. The red shaded areas represent the SD. The value in the lower right of each plot represents mean AUROC± SD.

**Figure 3: Predictive performance of intra-cohort four-fold cross-validation in MCO cohort and inter-cohort external validation in TCGA-CRC-DX cohort:** microsatellite instability (MSI), BRAF mutation status (BRAF), CpG island methylator phenotype (CIMP). **(a)** AUROC plots for four-fold cross-validation in MCO cohort. The red shaded areas represent the SD. The value in the lower right of each plot represents mean AUROC± SD. **(b)** AUROC plots for inter-cohort external validation in TCGA-CRC-DX cohort. The red shaded areas represent the 95% confidence interval (CI), which calculated by 1,000× bootstrap. The values in the lower right of each plot represents mean AUROC (95% CI).

**Figure 4. Test statistics for the pre-screening tool.** Test performance of MSI status, BRAF mutation and CIMP status in the TCGA-CRC-DX cohorts displayed as patients classified true/false positive/negative by the Swin-T model based on 95% sensitivity threshold, and fixed thresholds (0.25, 0.5 and 0.75).

**Figure 5. Predictive performance of Swin-T for prediction of MSI status using different**

**sizes of training data (MCO dataset).** The bar plot shows that the test performance (AUROC values) in external TCGA-CRC-DX dataset of Swin-T, when it is trained with different subset sizes of the MCO training dataset (25%,50%,75%, and 100%). The error bars represent the 95% confidence intervals.

**Figure 6. The visualization and interpretability of Swin-T model in predicting MSI status.** **(a)** Tile-level interpretation. we plot the top accurately predicted tiles from the top accurately scoring patients, i.e., for positive specimens (MSI-High), visualize the highest scoring tiles of the patients with the highest scores. Similar operation was done for negative samples (MSS), but the lowest scoring tiles are shown. 50 tiles from the MCO and TCGA datasets are visualized, separately. **(b)** Cell-level interpretation. we draw a cam map for a tile using Grad-CAM algorithm, which is a heatmap to highlight the areas (cells) that have greater attention weights within a high-resolution tile. The brighter the color for an area in the heatmap of a tile, the higher attention scores were assigned to this area by the model.

**Figure 1.**

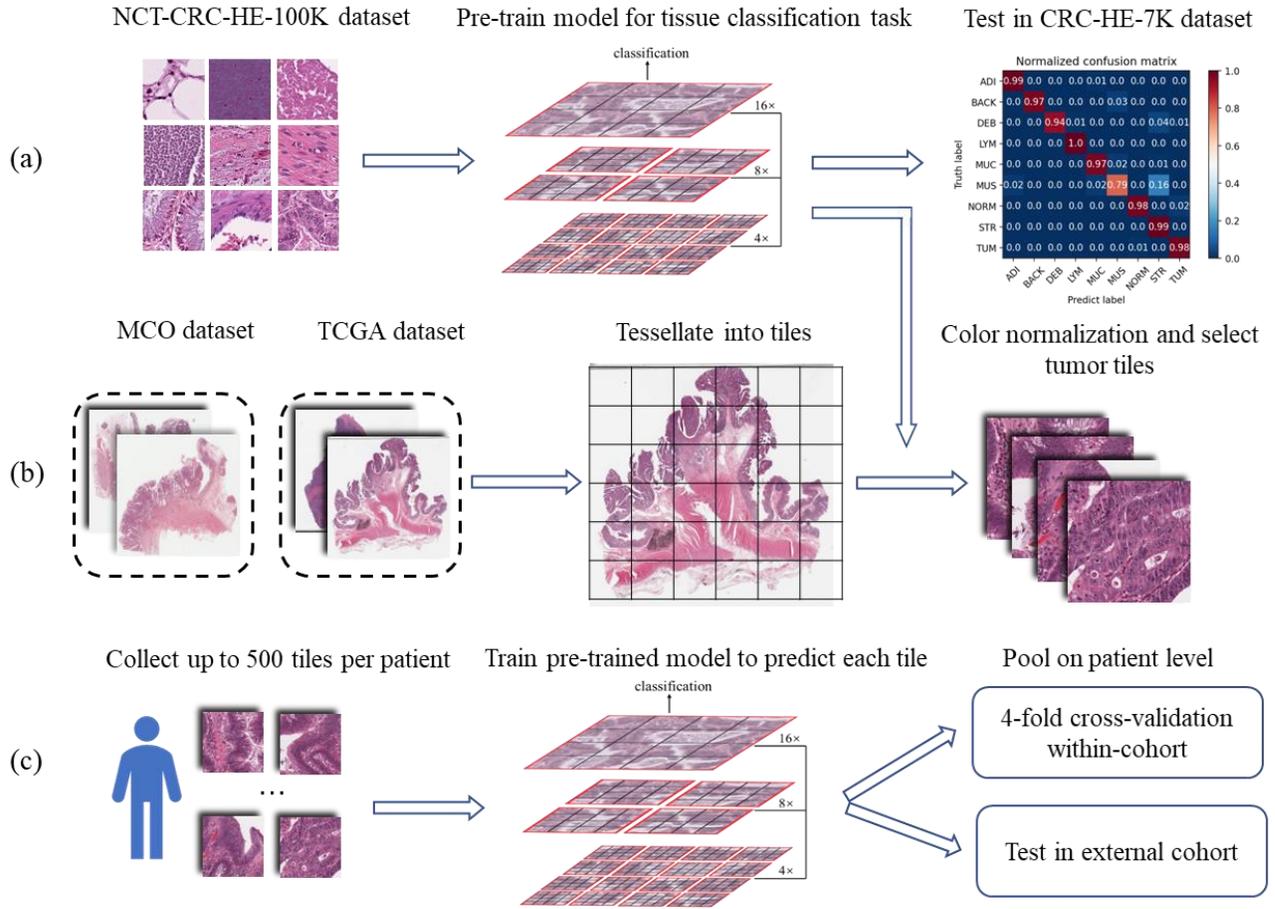

**Figure 2.**

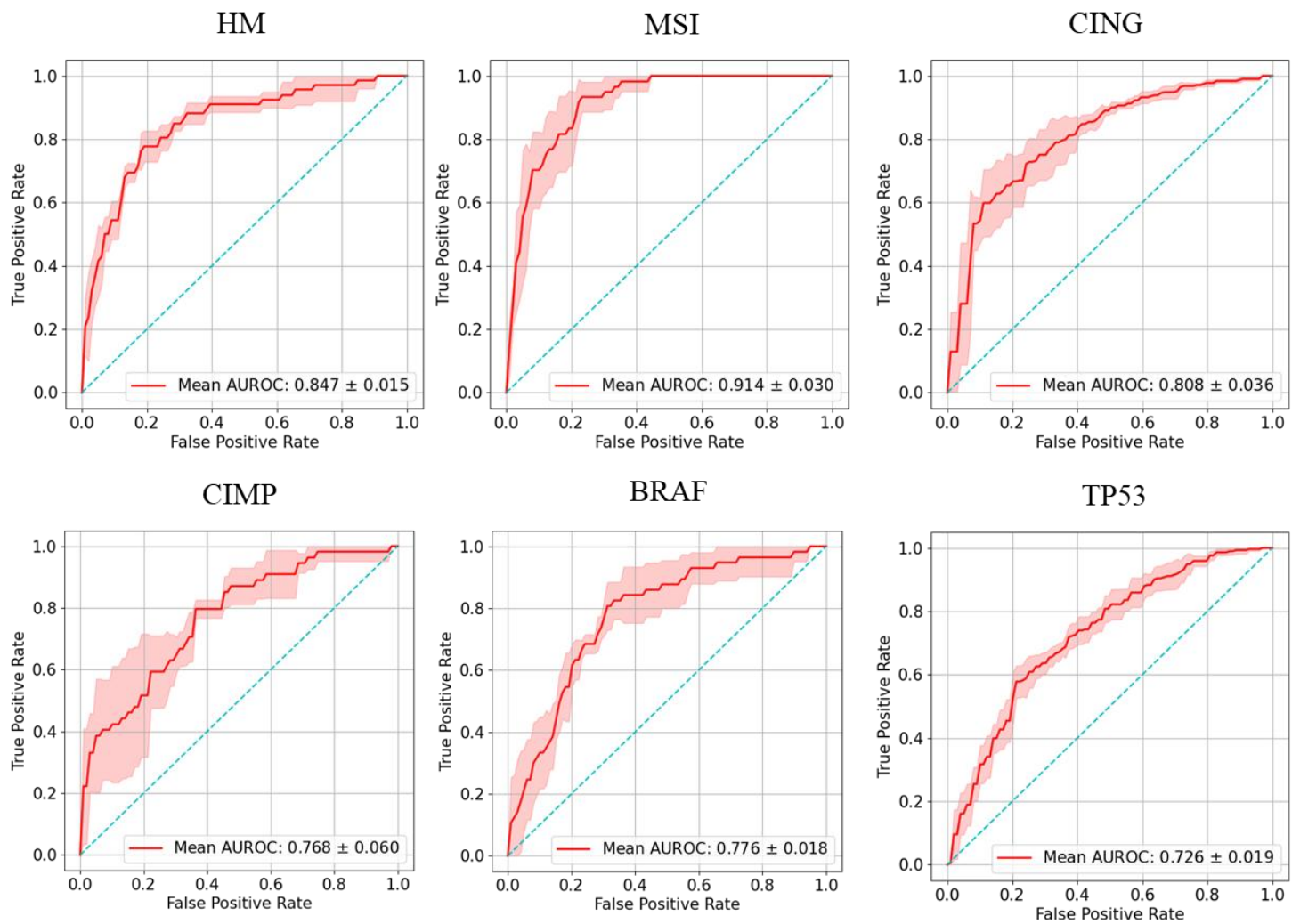

**Figure 3.**

## Intra-study cross validation in MCO

(a) 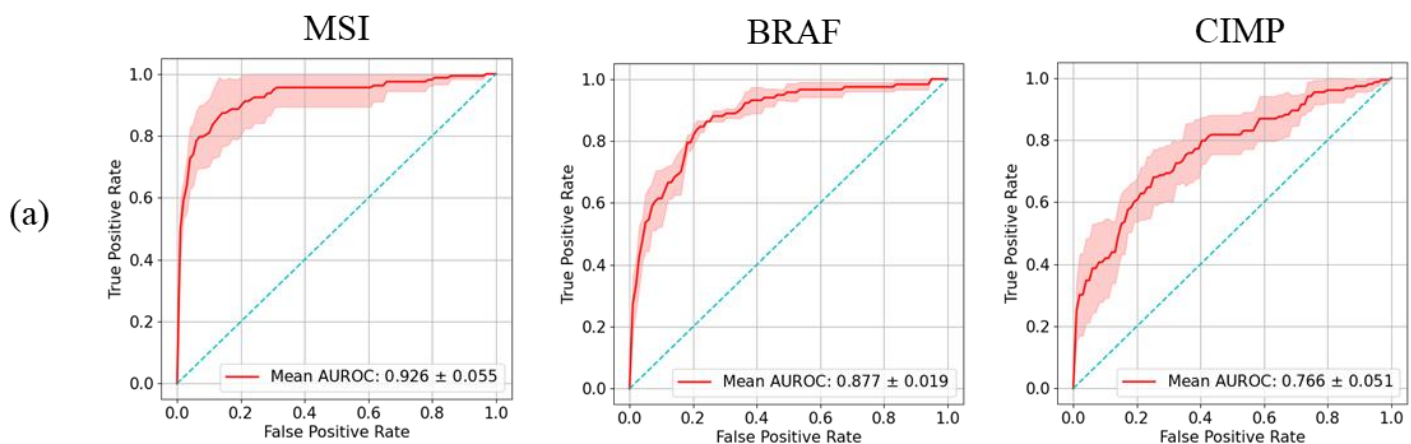

## Cross-study external validation in TCGA

(b) 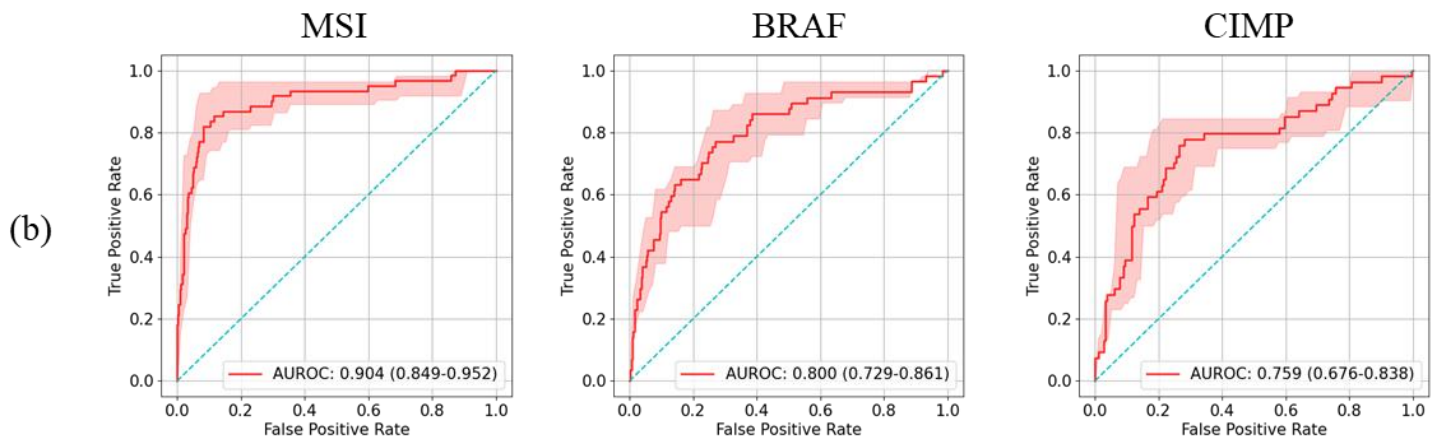

**Figure 4**

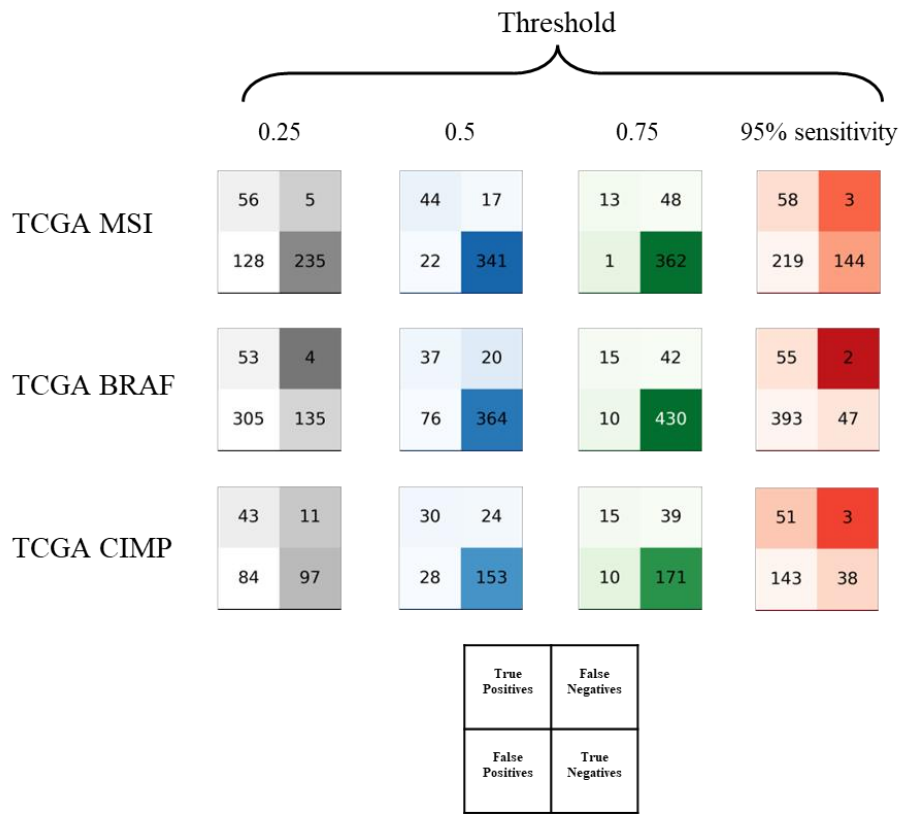

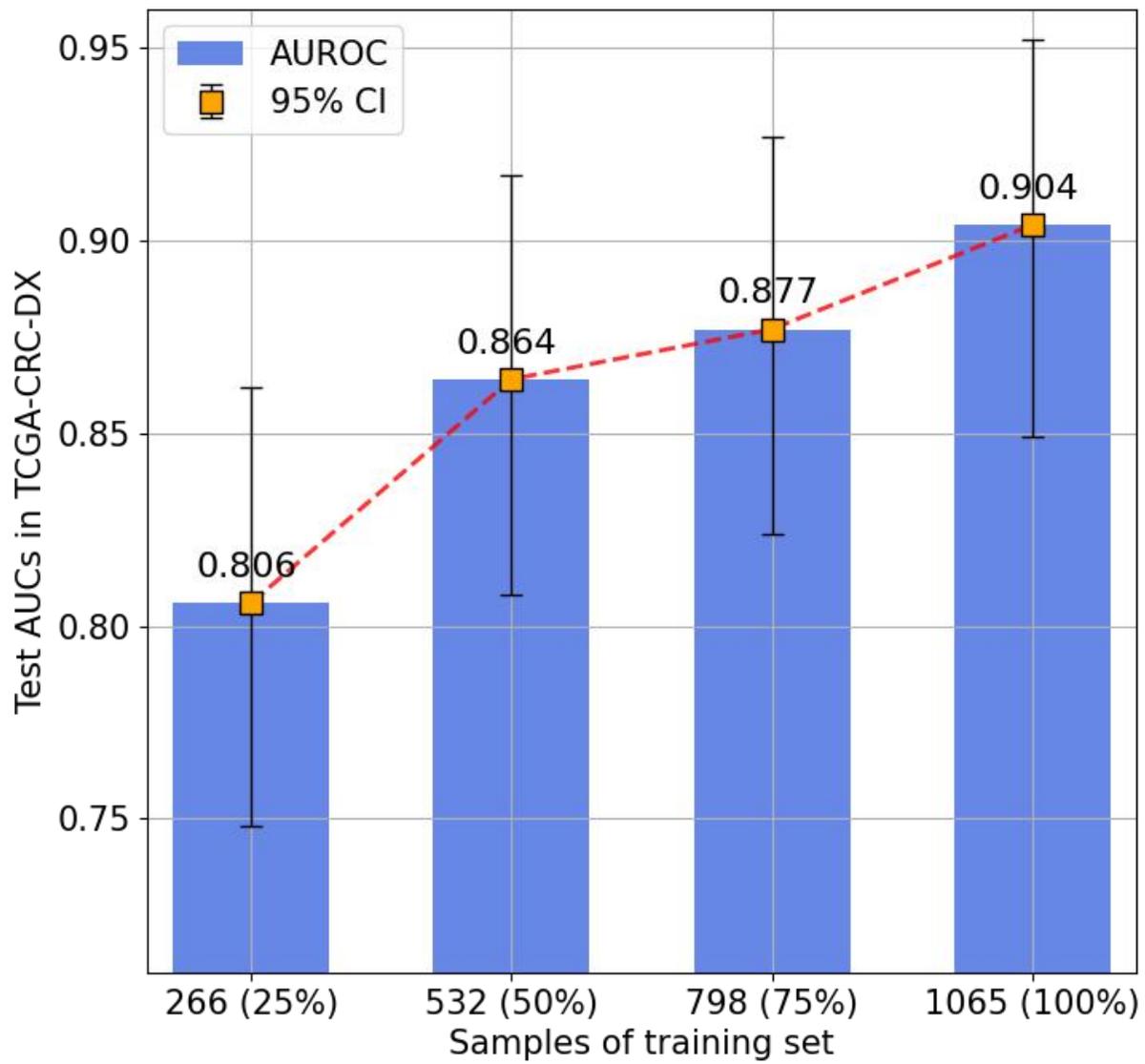

**Figure 5**

**Figure 6.**

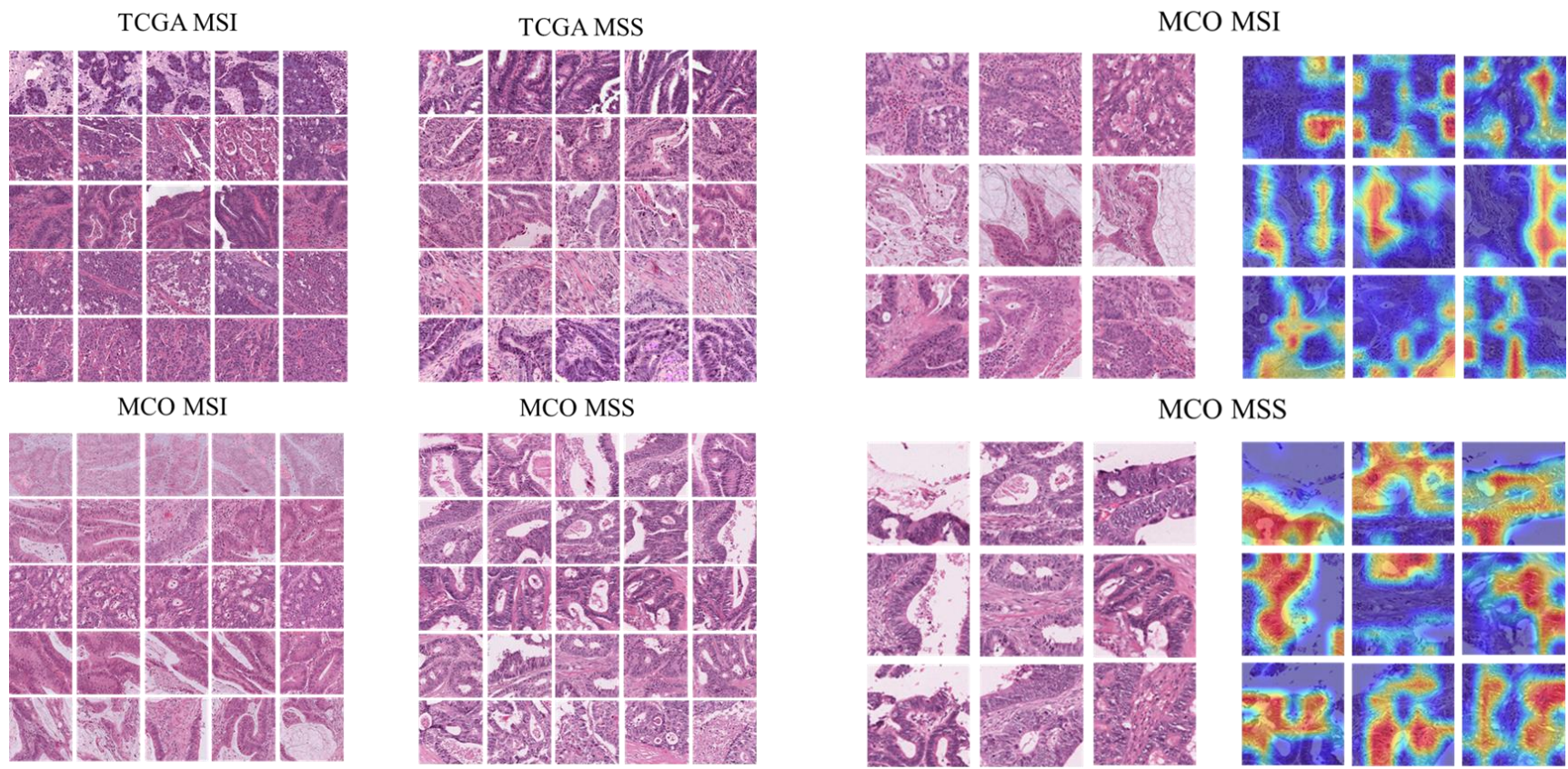

(a) (b)

# Supplement

**Supplement Figure 1: Results (AUPRC) of four-fold cross-validation of Swin-T based prediction of colorectal cancer pathways in the TCGA-CRC-DX cohort.** AUPRC plots for prediction of (a) hypermutation, (b) microsatellite instability, (c) chromosomal instability, (d) CpG island methylator phenotype, (e) BRAF mutation status and (f) TP53 mutation status. The red shaded areas represent the SD. The value in the lower right of each plot represents mean AUPRC ± SD.

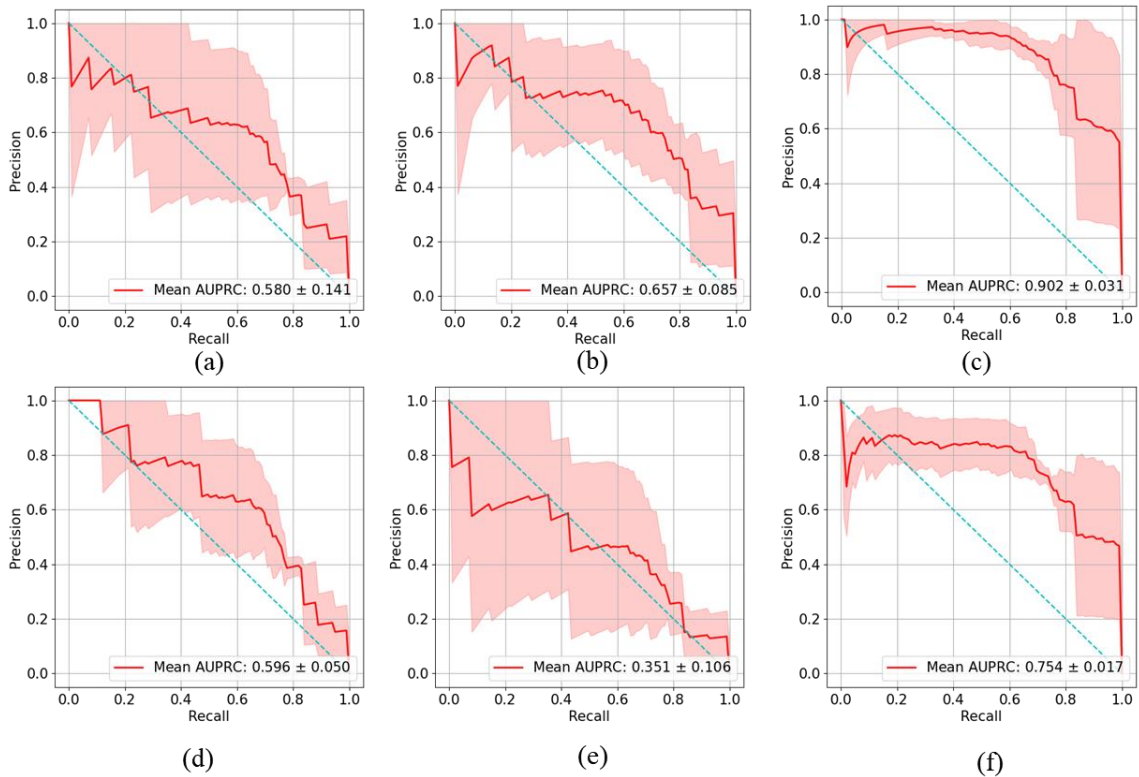